\documentclass{ws-procs9x6}
\usepackage{axodraw}
\begin{document}

\title{VARIOUS REALIZATIONS OF LEPTOGENESIS AND NEUTRINO MASS CONSTRAINTS
}

\author{T. HAMBYE
}

\address{Rudolf Peierls Center for Theoretical Physics \\
University of Oxford\\ 
1, Keble Road, 
Oxford OX1 3NP, UK\\
E-mail: hambye@thphys.ox.ac.uk}

\maketitle

\abstracts{Seven types of leptogenesis models which can lead to a 
successful explanation of baryogenesis are presented. Emphasis is put 
on the conditions which need to be fulfilled by the 
neutrino masses as well as by the heavy state masses. The model 
dependence of these conditions is discussed.}

\section{Introduction}

Following the recent convincing evidence for neutrino oscillations,
leptogenesis\cite{FY} 
has become a 
well motivated possible 
explanation of the origin of the baryon asymmetry of the universe.
In these proceedings, after a short introduction on the three basic 
ingredients of leptogenesis,
we will extend the discussion 
of Ref.\cite{MR} along two main directions. First, in the framework of 
the usual seesaw model 
with three right-handed 
neutrinos (``type-I'' seesaw model\cite{type-I}), we will study
the neutrino mass constraints, in particular the neutrino 
mass upper bound, which exist in order that leptogenesis 
can successfully explain the baryon asymmetry of the universe. 
We will show how this neutrino mass upper bound can be relaxed
by no longer assuming that the right-handed neutrinos 
have a hierarchical mass spectrum. 
Secondly, other models of leptogenesis which might also be attractive 
for various reasons will be introduced: ``type-II'' seesaw 
model with a scalar Higgs triplet, ``type-III'' seesaw model with 
fermion triplets, and combinations of them such as the
``type-I'' plus ``type-II'' model which is motivated by 
left-right or SO(10) models. The 
corresponding neutrino mass bounds will be discussed for all these models.
Finally, we will also briefly consider 
the case of radiative neutrino mass models.

\section{The Three Basic Ingredients of Leptogenesis}

There are essentially three ingredients to take 
into account for leptogenesis 
to work. In this section we will discuss only the case of 
the ``type-I`` seesaw model 
with three heavy right-handed neutrinos $N_i$
which is based on the following Lagrangian
\begin{equation}
\label{Lseesaw}
L  = L_{\rm SM} +\bar N_i i\partial\hspace{-1.3ex}/\, N_i +
(\lambda ^{ij} ~ H^\dagger N_i L_j   +
\frac{M_{N_i}}{2}   N_i N_i  +\hbox{h.c.})\,, 
\end{equation}
with $L_{j}= (\nu_{jL}$, $l_{jL})^T$, $H=(H^0,H^-)^T$. 
In Eq.~(\ref{Lseesaw})
we made the choice to work in the basis where the $N_i$ mass
matrix is real and diagonal. This model has 18 parameters; 9 combinations
of them enter in the neutrino mass 
matrix $M_{\nu}^I=-\lambda^T M_N^{-1} \lambda v^2$, and nine of them 
decouple from it. A very 
useful parametrization (in 
term of a orthogonal complex $R$ matrix) which allows to span easily 
all the parameter space of this model
in agreement with the data, and which is very useful 
to determine the neutrino mass bounds below, 
can be found in Ref.\cite{CI,HLNPS}. For the following we order 
the neutrino masses as: $m_{\nu_3}>m_{\nu_2}>m_{\nu_1}\ge0$ and  
$M_{N_3} \ge M_{N_2} \ge M_{N_1}\ge0$.

\subsection{The CP Asymmetry}

At a temperature well above their masses one can expect 
the right-handed neutrinos to be in thermal equilibrium 
in the universe thermal bath, 
due to Yukawa interactions or other possible interactions (such 
as gauge interactions in the right-handed sector). However, once 
the temperature drops below their masses the right-handed 
neutrinos disappear from the thermal bath by decaying to 
leptons and Higgs bosons.
The crucial quantity for leptogenesis is the CP-asymmetry, that is to say the 
averaged $\Delta L$ 
which is produced each time one $N_i$ decays. At lowest order, that is to say 
at one-loop order, for $N_1$ (and similarly for $N_{2,3}$), 
it is given by
\begin{equation}\label{eps}
\varepsilon_{N_1} \equiv \frac{\Gamma({N_1} \rightarrow l H^*) -
\Gamma( N_1 \rightarrow \bar{l} H)}
{\Gamma({N_1} \rightarrow l H^*) + \Gamma( N_1 \rightarrow \bar{l} H)}=
-\sum_{j=2,3}\frac{3}{2}
  \frac{M_{N_1} }{M_{N_j} }\frac{\Gamma_{N_j} }{M_{N_j} }
  I_j\frac{2 S_j + V_j}{3} 
\end{equation}
where
\begin{equation}\label{IGamma}
I_j = \frac{ \hbox{Im}\,[ (\lambda  \lambda ^\dagger)_{1j}^2 ]}
{|\lambda \lambda ^\dagger |_{11} |\lambda \lambda ^\dagger |_{jj}}
 \, ,\qquad
\frac{\Gamma_{N_j}}{M_{N_j}} = \frac{|\lambda \lambda ^\dagger |_{jj}}{8\pi}
\equiv \frac{\tilde{m}_j M_{N_j}}{8\pi v^2}
\,,
\end{equation}
and where
\begin{equation}
S_j = \frac{M^2_{N_j}  \Delta M^2_{1j}}{(\Delta M^2_{1j})^2+M_{N_1} ^2
   \Gamma_{N_j} ^2} \,, \,\,\,\,\,
V_j = 2 \frac{M^2_{N_j} }{M^2_{N_1}}
\bigg[ \big(1+\frac{M^2_{N_j} }{M^2_{N_1}}\big)\log\big(1+
\frac{M^2_{N_1}}{M^2_{N_j} }\big)
- 1 \bigg],
\end{equation}
with $\Delta M^2_{ij}=M^2_{N_j}-M^2_{N_i}$.
The factors $S_j$ ($V_j$) comes from the one-loop 
self-energy (vertex) contribution to the 
decay widths, Fig.~1. 
The $I_j$ factors are the CP-violating coupling combinations entering 
in the asymmetry.
%
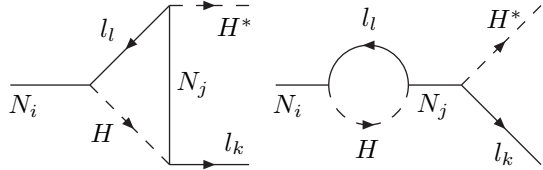
\begin{figure}[b]
\begin{center}
\begin{picture}(200,60)(0,0)
\Line(0,30)(30,30)
\DashArrowLine(60,60)(90,60){5}
\ArrowLine(60,0)(90,0)
\Line(60,60)(60,0)
\ArrowLine(60,60)(30,30)
\DashArrowLine(30,30)(60,00){5}
\Text(5,22)[]{$N_i$}
\Text(35,12)[]{$H$}
\Text(37,51)[]{$ l_l$}
\Text(69,30)[]{$N_j$}
\Text(85,52)[]{$H^\ast$}
\Text(85,8)[]{$ l_{k}$}
\Line(100,30)(120,30)
\DashArrowArc(135,30)(15,180,360){5}
\ArrowArc(135,30)(15,0,180)
\Line(150,30)(170,30)
\DashArrowLine(170,30)(200,60){5}
\ArrowLine(170,30)(200,0)
\Text(105,22)[]{$N_i$}
\Text(135,6)[]{$H$}
\Text(136,55)[]{$ l_l$}
\Text(160,22)[]{$N_j$}
\Text(187,55)[]{$H^\ast$}
\Text(187,4)[]{$ l_{k}$}
\end{picture}
\end{center}
\caption{One-loop diagrams contributing to the asymmetry
from the $N_i$ decay.}
\label{fig1}
\end{figure}

\subsection{The Efficiency Factor}

Once the averaged $\Delta L$ produced per decay has been calculated, the 
second ingredient to consider 
is the efficiency factor $\eta$. This factor allows to 
calculate the lepton asymmetry produced from the CP-asymmetry,
\begin{equation}
\frac{n_L}{s}=\varepsilon_{N_i} Y_{N_i}|_{T >> M_{N_i}} \eta \,,
\end{equation}
where $Y_{N_i}=n_{N_i}/s$ is the number density 
of $N_i$ over the entropy density,
with $Y_{N_i}|_{T >> M_{N_i}}=135 \zeta(3)/(4 \pi^4 g_*)$ 
where $g_*=112$ is the number of degrees of freedom 
in thermal equilibrium in the ``type-I'' model before the $N_i$ decayed. If 
all right-handed neutrinos
decay out-of-equilibrium, the lepton asymmetry produced is 
just given by the CP asymmetry times the number of $N_i$ over the entropy 
density before the $N_i$ decayed, i.e.~$\eta=1$. 
However, the efficiency 
factor can be much smaller than one, if
they are not fully out-of-equilibrium while decaying, 
and/or if there are at this epoch 
L-violating processes partly in thermal equilibrium.
The processes which can put the $N_i$ in thermal equilibrium and/or violate 
L are the inverse decay process and $\Delta L=1,2$ scatterings. 
To avoid a large damping effect, it is necessary 
that these processes are not too fast with respect to the Hubble 
constant. For the 
inverse decay process (which is the 
most dangerous process, see e.g. the discussion of Ref.\cite{DiBari}),
this gives the 
condition: $\Gamma_{N_i}/H(T \simeq M_{N_i}) \le 1$ 
with $H(T)=\sqrt{4 \pi^3 g_* / 45} \, T^2/M_{\hbox{\footnotesize Planck}}$.
In practice to calculate $\eta$ we need to put all these processes
in the Boltzmann equations\cite{plum,GNRRS} which allow a precise
calculation of the produced lepton asymmetry
as a function of the temperature $T$. The corresponding efficiency factor
including finite temperature effects can be found in Ref.\cite{GNRRS} 
in the limit where 
the right-handed neutrinos have a 
hierarchical spectrum $M_{N_1} << M_{N_{2,3}}$. In this limit
only the 
asymmetry produced by the decay of the lightest 
right-handed neutrino $N_1$ survives 
and is important, which simplifies greatly the calculations.

\subsection{The $B$ to $L$ Asymmetry Conversion}

Once the $L$ asymmetry $n_L/s$ has been produced and determined, 
the last step is to calculate the corresponding baryon asymmetry 
produced due to the partial conversion of the $L$ 
asymmetry to a $B$ asymmetry by 
the Standard Model non-perturbative sphaleron processes.
The conversion factor is $n_B/s=-(28/79) n_L/s$.\cite{conversfactor}
It comes from taking into account the fact 
that these processes conserve $B-L$, violate 
$B+L$ and are  
fast (i.e.~in thermal equilibrium) above the electroweak scale.
From this conversion factor we finally get the total baryon 
asymmetry produced $n_B/s=-1.38 \cdot 10^{-3} \varepsilon_{N_1} \eta$ 
which has to be compared 
with the experimental BBN and WMAP value for this 
ratio: $n_B/s=(8.7 \pm 0.4 ) \cdot 10^{-11}$.

\section{The Neutrino Mass Constraints}

In addition to the three ingredients above there is a fourth crucial 
ingredient which makes the all interest of the leptogenesis mechanism: 
the neutrino mass constraints. These constraints come from the fact 
these are the same interactions 
which produce both the neutrino masses and leptogenesis.
There are two types of neutrino mass constraints on leptogenesis. 

\subsection{The Neutrino Mass Constraint on the Efficiency}

The first constraint
is the neutrino mass constraint on the 
size of the washout, that is to say on $\eta$.
It comes from the fact that, in full generality,
the ratio $\Gamma_{N_i}/H$, which has to be smaller than unity to have 
no washout suppression, is always larger
than the ratio of the lightest neutrino mass $m_{\nu_1}$ over 
the $m_*= 16 \pi^2 v^2 
\sqrt{g_* \pi/45}/M_{\hbox{\footnotesize Planck}}\,$$\sim 10^{-3}$~eV 
scale:\cite{BP2}
\begin{equation}
\frac{\Gamma_{N_i}}{H} \ge \frac{m_{\nu_1}}{10^{-3}\,\hbox{eV}}\,.
\label{outofeq}
\end{equation}
This inequality comes simply from Eq.~(\ref{IGamma}) 
and the $v^2 | \lambda \lambda^\dagger |^2_{jj} / M_{N_j} \ge m_{\nu_1}$ 
inequality. It 
means that, if $m_{\nu_1}> 10^{-3}$~eV, there will 
be some washout and the larger is $m_{\nu_1}$ above this 
scale the larger is the washout, i.e.~the smaller is $\eta$.
For example for $m_{\nu_1}=\sqrt{\delta 
m^2_{\hbox{\footnotesize atm}}} \simeq 0.05$~eV 
the value of $\eta$ is $\sim 0.01$ for 
$M_{N_1} \le 10^{14}$~GeV. For 
$m_{\nu_1}=0.5$~eV this value becomes few $10^{-4}$ for 
$M_{N_1} \le 10^{12}$~GeV and decreases for larger values of $M_{N_1}$.
The fact that $m_*$, which is a function of the electroweak and Planck 
scales, is of order the neutrino 
masses is a quite remarkable fact since it means that for leptogenesis 
the $N_i$ are naturally only slightly in thermal equilibrium or 
out-of-equilibrium.
Note that Eq.~(\ref{outofeq}) is often rewritten in term of the so-called 
effective masses of Eq.~(\ref{IGamma}) 
as $\tilde{m}_i \ge 10^{-3}$~eV.
 
\subsection{The Neutrino Mass Constraint on the Size of $\varepsilon_{N_i}$}

The second neutrino mass constraint is on the size of 
the CP-asymmetries $\varepsilon_{N_i}$.
It generally applies but, contrary to the first constraint above, it 
do not always applies.
This depends on the type of mass spectrum the right-handed 
neutrino have: very hierarchical, ``normally'' 
hierarchical or quasi-degenerate.

\subsubsection{Very hierarchical right-handed 
neutrinos: $M_{N_1} << M_{N_{2,3}}$}

If right-handed neutrino masses differ by several 
orders of magnitude, the size of the 
$\varepsilon_{N_1}$ asymmetry is quite 
constrained by the size of the neutrino masses. There exists an upper 
bound\cite{epsbound,TH,exactepsbound} 
which in its exact form was given by Ref.\cite{exactepsbound}:
\begin{equation}
\label{epsDI}  
|\varepsilon_{N_1}| \le 
  \frac{3}{16\pi}\frac{M_{N_1}}{v^2}(m_{\nu_3} - m_{\nu_1})=
\frac{3}{16\pi}\frac{M_{N_1}}{v^2}
\frac{\Delta m^2_{\hbox{\footnotesize atm}}}{m_{\nu_3} + m_{\nu_1}}\,. 
\end{equation}  
Since $\Delta m^2_{\hbox{\footnotesize atm}}\simeq 2 \cdot 10^{-3}$~$eV^2$ is 
fixed experimentally, this bound 
decreases as the neutrino masses increase.
Therefore as the neutrino masses increase there are two suppression 
effects arising: the washout effect increases and the upper bound on the 
asymmetry decreases. Successful leptogenesis leads therefore to the upper 
bound:\cite{epsP,GNRRS,HLNPS}
\begin{equation}
\label{mnu3bound}
m_{\nu_3} < 0.12-0.15~\hbox{eV} \,.
\end{equation}
For more details see Ref.\cite{MR} in these proceedings.
Note that in order to derive this bound Eq.~(\ref{epsDI}) is not sufficient. 
As the final produced lepton asymmetry depends not only on $m_{\nu_3}$ 
and $M_{N_1}$ but also, through 
the efficiency factor, 
on $\Gamma_{N_1}$ (or $\tilde{m}_1$),
it is necessary\cite{epsP,HLNPS} to have an upper 
bound 
for fixed values of these 3 parameters and not only as a 
function of $m_{\nu_3}$ and $M_{N_1}$ as in Eq.~(\ref{epsDI}).
This bound can be found in Ref.\cite{HLNPS}.
Note also that, due to the fact that
the upper bound on the CP-asymmetry 
is proportional to $M_{N_1}$, successful leptogenesis with 
hierarchical right-handed neutrinos 
implies also a lower bound on this 
mass:\cite{exactepsbound,epsbound,TH,GNRRS}
\begin{equation}
\label{MN1bound}
M_{N_1} > 5 \times 10^{8}\,\hbox{GeV} \,.
\end{equation}
This result holds for the case where  the $N_1$ are in thermal equilibrium 
before decaying. Starting instead (due to inflation dynamics) 
from a universe with no (with only) right-handed neutrinos at a 
temperature above their mass, this bound becomes\cite{GNRRS}: $M_{N_1} > 2 
\times 10^9$~GeV, ($2 \times 10^7$~GeV).

\subsubsection{``Normally'' hierarchical right-handed neutrinos}

If right-handed neutrinos have a hierarchy similar to the ones of the 
charged leptons or quarks, that is to say if $M_{N_1} \simeq (10-100) M_{N_2}$
with $M_{N_3} > M_{N_2}$, the L-asymmetry production is still dominated by 
the decays of the lightest right-handed neutrino $N_1$. In this case the 
upper bound on 
the CP-asymmetry is the same as for very-hierarchical neutrinos, except 
that there are extra corrections\cite{HLNPS} 
in $M_{N_1}^2/M_{N_{2,3}}^2$ to be added 
in Eq.~(\ref{epsDI}):
\begin{equation}
\delta\varepsilon_1 \simeq 
  \frac{3}{16\pi}\frac{M_{N_1}}{v^2} \tilde{m}_{2,3}
\frac{M_{N_1}^2}{M_{N_{2,3}}^2} \,.
\end{equation}
These corrections generically are small so that all the bounds obtained 
in the previous section are still valid, but not always.
There exist configurations of the Yukawa couplings which lead to 
large corrections. 
The point is that, contrary to the leading term in 
Eq.~(\ref{epsDI}), the corrections do not decrease 
when the neutrino masses decrease and they do not necessarily vanish 
for degenerate light-neutrino masses.
As a result\cite{HLNPS}, for special
configurations giving large $\tilde{m}_{2,3}$ 
but small neutrino masses, one can have successful leptogenesis 
with $M_{N_1}$ well below the lower bound of Eq.~(\ref{MN1bound})
and with neutrino masses well above the bound of Eq.~(\ref{mnu3bound}).
An explicit example of such  Yukawa coupling configuration leading to 
successful leptogenesis with $M_{N_1} \simeq 10^6$~GeV has 
been recently considered in Ref.\cite{RST}.  

\subsubsection{Quasi-degenerate right-handed neutrinos}

If at least two right-handed neutrinos have masses very close to each other,
$M_{N_1} \sim M_{N_2}$, the situation is changing drastically with respect 
to the two previous cases.
This is due to the fact that the one loop self-energy diagram of Fig.1 
displays
in this case
a resonance behaviour\cite{flanz,HLNPS,HMW} coming 
from the propagator of the virtual 
right-handed neutrino in this diagram. This effect can be seen from the $S_j$ 
factors in Eq.~(\ref{eps}).
Since in the seesaw model the decay widths of the $N_i$ are generically 
much 
smaller than their masses, this resonance effect can lead to a 
several order of magnitude enhancement of the asymmetry.
At the resonance, that is to say for $M_{N_2}-M_{N_1}=\Gamma_{N_2}/2$, 
the $S_{2}$ factor, which is unity in the 
$M_{N_1} << M_{N_{2,3}}$ limit, is as large as 
$\frac{1}{2}M_{N_2}/\Gamma_{N_2}$.
In this case, up to a $1/2$ factor, the $S_2$ factor 
compensates the $\Gamma_{N_2}/M_{N_2}$ prefactor in Eq.~(\ref{eps}) which 
gives $\varepsilon_{N_1}=\frac{1}{2} I_2$.
Moreover in this case one can show that the asymmetry is not 
bounded anymore by an expression depending on the neutrino masses. In
particular the upper bound is not proportional to 
$\Delta m^2_{\hbox{\footnotesize atm}}/(m_{\nu_3}+ m_{\nu_1})$ as 
in the hierarchical 
case,\footnote{One can check\cite{exactepsbound,HLNPS} that this neutrino 
mass factor comes from the fact that 
for hierarchical right-handed neutrinos $S_2=S_3$. This equality 
has no reason to be true anymore with 
quasi-degenerate right-handed neutrinos.} Eq.~(\ref{epsDI}).
The Yukawa coupling factor $I_2$ turns out to be bounded just by unity 
so that the upper bound on $\varepsilon_{N_1}$ is just $\frac{1}{2}$ 
independently of light and heavy neutrino masses\cite{HLNPS}.
Together with $\varepsilon_{N_2}$, which is equal to $\varepsilon_{N_1}$ 
in this case and has also to be taken into account, 
CP-violation is just bounded by unity.
As a result, since neither the maximal asymmetry nor the washout 
effect (coming from inverse decay\footnote{Note that there is 
a $M_{N_i}$ mass dependence in the washout coming from $\Delta L =2$ 
scatterings. This effect however can be large only for large 
values of the $M_{N_i}$ close to the GUT scale.}) depend on $M_{N_1}$, 
successful leptogenesis can be obtained at any scale except 
that the $L$ to $B$ conversion from sphalerons still needs to be 
effective. This requires $M_{N_1}$ to be above the electroweak 
scale (i.e.~typically above $\sim 1$~ TeV).
Moreover since the upper bound on the CP asymmetry
is independent of neutrino masses, there is no more suppression of 
the asymmetry for large neutrino masses. The only remaining suppression 
effect arising for large neutrino masses is the one of section 3.1 above 
coming 
from the washout. 
Therefore the upper bound on the neutrino masses in this case gets 
considerably relaxed. This is shown in Fig.2.a where is plotted, as a 
function of $m_{\nu_3}$, the level of degeneracy which is needed to 
have successful leptogenesis. Values far above the eV are possible 
which means that {\it in full generality there is no more relevant upper 
bound on neutrino masses coming from leptogenesis}.
The value $m_{\nu_3} \simeq 1$~eV can lead to successful leptogenesis
with a level of degeneracy of 
order $(M_{N_2}-M_{N_1})/M_{N_2}\simeq 4 \cdot 10^{-2}$ which 
is quite moderate.

\begin{figure}[t]
\centerline{\epsfxsize=2.0in{\epsfbox{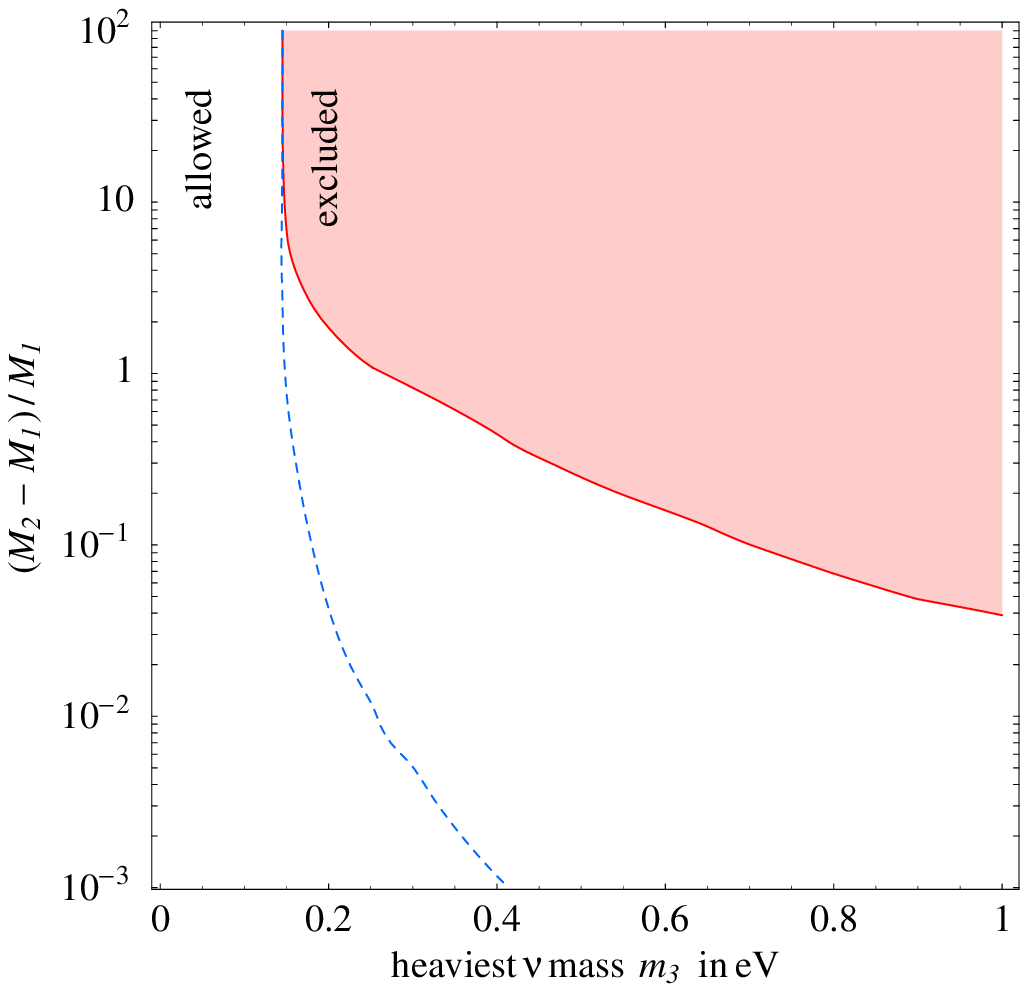}} \hspace{0.1cm}
\epsfxsize=2.0in{\epsfbox{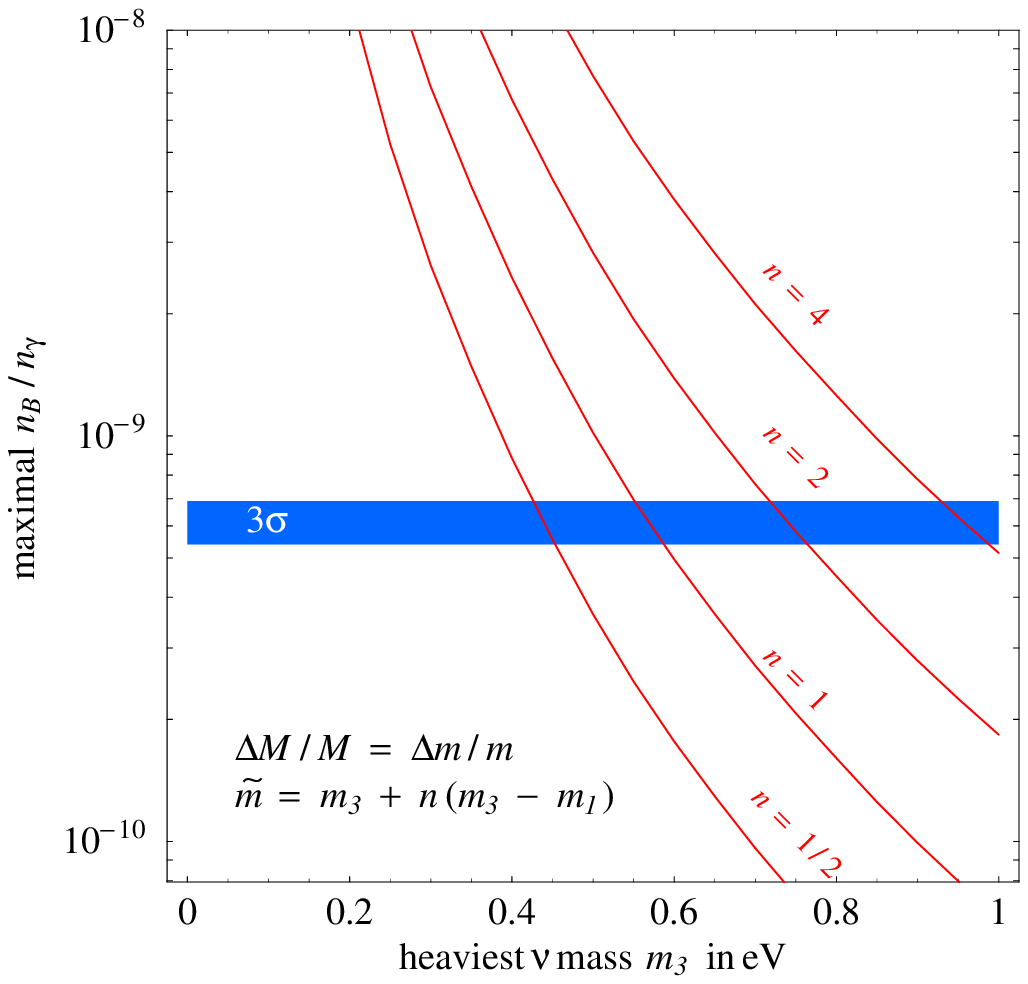}}} 
\caption{Fig.2.a shows$^5$
how much the
maximal value of neutrino mass compatible with thermal
leptogenesis increases
when right-handed neutrinos are allowed to be
quasi-degenerate. The dashed line is what we obtain if we take into 
account only the resonance enhancement effect, not the neutrino 
mass effect,
see text and Ref.$^5$.
Fig.1.b holds assuming Eqs.~(\ref{splitsol})-(\ref{splitatm}) 
as a generic example 
that everything is as degenerate as neutrinos,
considering
the natural possibility that the $\tilde{m}_i$ should be
quasi-degenerate with the neutrino masses, taking
$\tilde{m}_{1,2,3} < m_{\nu_1}+n(m_{\nu_3}-m_{\nu_1})$. 
}
\vspace*{-2mm}
\end{figure}

One could argue that such a degeneracy of right-handed neutrinos
is unnatural and that we would 
anyway generally expect a hierarchical spectrum for the right-handed neutrino
by analogy with charged leptons or quarks.
However, since the bound on neutrino masses from leptogenesis is relevant
  only for a quasi-degenerate spectrum of light neutrinos, when
  calculating this bound one can
  wonder what would be the most natural right-handed neutrino mass
  spectrum to explain such correlations between the light-neutrino
  masses.
Presumably, three
quasi-degenerate right-handed neutrinos are
the most natural spectrum in this case (with Yukawa couplings
correlated among each other). Such a quasi degenerate spectrum, 
unlike a hierarchical spectrum, wouldn't require in addition precise
correlations between the Yukawa couplings and the right-handed
neutrino masses (in order that the ratio 
between the $M_{N_i}$ is compensated in the seesaw mass 
formula by the same ratio
between the Yukawa couplings). Therefore 
the bound given in Fig.~2.a is the fully relevant one.

To complete this section note  
that there exists nevertheless one specific pattern,
which is probably the most realistic one, which leads
to more stringent bounds.\cite{HLNPS}
In fact, if neutrinos were quasi-degenerate,
the degeneracy would presumably not be accidental but due to
some reason:
a broken SO(3) flavour symmetry is probably the simplest possibility.
One expects that in such framework all quantities, and not only
neutrino masses, are close to the ideal limit
where three degenerate right-handed neutrinos
give equal masses to three orthogonal combinations of left-handed  
neutrinos.
Therefore one expects something 
like $\tilde{m}_i - \tilde{m}_j\approx m_{\nu_i} - m_{\nu_j}$ and
\begin{eqnarray}
&&\frac{M_{N_2}-M_{N_1}}{M_{N_1}}\sim \frac{m_{\nu_2}-m_{\nu_1}}{m_1}\approx
\frac{\tilde{m}_2-\tilde{m}_1}{\tilde{m_{\nu_1}}}
 \approx \frac{\Delta m^2_{\rm sun}}{2 m_{\nu_1}^2}\approx
0.5~10^{-4} \bigg(\frac{\hbox{eV}}{m_{\nu_2}}\bigg)^2 \label{splitsol}
 \,\,\,\,\,\,\,\,{}    \\
&&\frac{M_{N_3}-M_{N_2}}{M_{N_2}}\sim \frac{m_{\nu_3}-m_{\nu_2}}{m_{\nu_2}} 
\approx
\frac{\tilde{m}_3-\tilde{m}_2}{\tilde{m_2}}
\approx \frac{\Delta m^2_{\hbox{\footnotesize atm}}}{2 m_{\nu_3}^2}\approx
10^{-3} \bigg(\frac{\hbox{eV}}{m_{\nu_2}}\bigg)^2.\,\,\,\,\,{} 
\label{splitatm}
\end{eqnarray}
Right-handed neutrinos can be more degenerate than in the above  
estimates
if only the neutrino Yukawa couplings
deviate from the symmetric limit, and can be less degenerate only
if there are accidental cancellations between non-universal
Yukawa couplings and non-degenerate $M_{N_{1,2,3}}$
in the see-saw prediction for neutrino masses.

Assuming the relations of Eqs.~(\ref{splitsol})-(\ref{splitatm}),
the bound on the neutrino masses can be read off from Fig.~2.b
where we give the maximal
baryon asymmetry we obtain as a function of $m_{\nu_3}$ for values of
$\tilde{m}_{1,2,3} < m_{\nu_1}+n(m_{\nu_3}-m_{\nu_1})$ 
with $n=\{1/2,1,2,4\}$. Taking  
$\tilde{m}_{1,2,3} < m_{\nu_3}$ ($n=1$),
as the generic example for the case that the $\tilde{m}$ would be
precisely of order the neutrino masses, gives the constraint
\begin{equation}
m_{\nu_3} < 0.6 \,\, \hbox{eV} \,,  \label{m3reali}
\end{equation}
which is stronger than in the fully general case of Fig.~2.a.
The suppression effect comes essentially from the fact that for values 
of $\tilde{m}_{1,2,3}$ close to the neutrino masses the Yukawa 
coupling factors $I_j$ are suppressed\cite{HLNPS,epsP}.
The result is quite sensitive to how these quantities are close to each 
other. For example taking n=4 (which could be considered as a quite moderately
fine-tuned case) leads already to an upper bound as large as 1 eV.

In summary without a predictive flavour model which would show how
the correlations between the seesaw parameters
at the origin of the degenerate spectrum occur,
in order to have a safe bound we must consider the fully general case
of Fig.~2.a
 (where $n$ was left as a free parameter in order to maximize the asymmetry).
Even in a very constrained situation the
neutrino masses
can be as large as 0.6~eV, Eq.~(\ref{m3reali}), or 1~eV.\footnote{Stronger 
constraints will arise if supersymmetry exists
and if right-handed neutrinos lighter than $10^{10}$~GeV
will be needed to avoid gravitino overproduction.}

\section{Leptogenesis in the Framework of Other Seesaw Models}

The type-I seesaw mechanism is probably the most
direct extension of the Standard Model 
we can consider in order to explain the neutrino 
masses and is in this sense the most attractive. However it is not the only 
attractive seesaw model.
Beside the type I seesaw, one can think about 
two other basic seesaw mechanisms. The first one is the type-II seesaw 
model\cite{type-II}
where 
neutrino masses are due to the exchange of an heavy scalar Higgs triplet.
The second one is from the exchange of three heavy self-conjugated 
$SU(2)_L$ triplets of fermions\cite{type-III}, a 
model which according to us should be 
called type-III seesaw model as it induces
the neutrino mass from a third type of heavy particles.
In addition to these three seesaw basic mechanisms one can also think about 
combinations of them. In the following we 
consider these various alternatives and see whether they can lead 
to successful leptogenesis and what are the corresponding mass bounds.

\subsection{The Type-II Seesaw Model}

The type-II seesaw model with just one heavy scalar triplet $\Delta_L$ which 
couples to 2 leptons doublets and to two Higgs doublets is 
a particularly minimal model. It is based on the Lagrangian
\begin{equation}
\label{Lseesaw2}
 L \owns 
-M^2_\Delta Tr\Delta_L^\dagger \Delta_L
 - (Y_\Delta)_{ij} L_{i}^T C i \tau_2 
\Delta_L L_{j} + \mu H^T i \tau_2 \Delta_L H + h.c. \,,
\end{equation}
with 
\begin{equation}
\Delta_L=
\begin{pmatrix}
\frac{1}{\sqrt{2}}\delta^+ & \delta^{++}  \\
\delta^0 & - \frac{1}{\sqrt{2}} \delta^+ 
\end{pmatrix} \,.
\end{equation}
It leads to the neutrino mass matrix: $M_\nu^{II} = 2 Y_\Delta 
v_\Delta \simeq 2 Y_\Delta \mu^* v^2/M^2_\Delta$.
This model in full generality has only 11 parameters, the 
triplet mass, its $\mu$ coupling and 6 real parameters plus 
3 phases 
in the $Y_\Delta$ Yukawa coupling matrix. 
This model has the attractive property that the knowledge of 
the full low energy neutrino mass matrix $M_\nu$ 
would allow to determine the full 
flavour high energy structure in $Y_\Delta$, both matrices being just 
proportional to each other.
However for leptogenesis it turns out that this model is too minimal.
As explained in Ref.\cite{MS1,HMS1,HLNPS,DHHRR}, 
since the triplet is not 
a self-conjugated 
particle, there is no vertex diagram and leptogenesis could 
come only from a self-energy diagram involving two leptons in the final state
and two Higgs doublets in the self-energy, third diagram of Fig.3. 
This diagram with just one triplet
is real and therefore doesn't bring any CP-violation. At two loops the 
asymmetries are too suppressed.
Therefore the  standard model 
extended by just one scalar triplet leads in a ``minimal'' 
way to neutrino masses but do not lead to 
successful leptogenesis.

However, based on the type-II model with 
just one triplet coupling to leptons, there is one framework which
can work. It is the supersymmetric version of this model, i.e.~the MSSM 
model extended by 
a pair of scalar triplets with opposite hypercharges.
In a way similar to the soft leptogenesis mechanism with right-handed 
neutrinos\cite{SOFTN,MR}, the 
supersymmetry breaking terms involving the triplets
can remove the mass degeneracy between the two triplets and lead to 
resonant 
leptogenesis from the self-energy diagram involving the two 
triplets\cite{DHHRR}.
This requires triplet mass between $\sim 10^{3}$~GeV and 
$\sim 10^{9}$~GeV.\footnote{Note that triplets, unlike right-handed 
neutrinos have gauge scatterings, which tend to put them in closer 
thermal equilibrium, reducing the efficiency. The exact 
efficiency for a scalar triplet is yet to be calculated but the one for 
a fermion triplet is known\cite{HLNPS}. All lower bounds 
on the scalar triplet masses we give here are 
obtained using the 
fermion triplet efficiency, assuming that both efficiencies are same, as
gauge scatterings are expected to be similar up to factors of order one.
This should be checked explicitly.}

\subsection{The Heavy Triplet of Fermion Model}

The third seesaw basic way to induce neutrino masses is by adding to 
the Standard Model 3 self-conjugated $SU(2)_L$ triplets of 
fermions.\cite{type-III}
The Lagrangian keeps the same structure as the one of the type-I seesaw 
model but with different $SU(2)_L$ contractions:
\begin{equation}
L  = L_{\rm SM} +\bar N^a_i iD\hspace{-1.3ex}/\, N^a_i +
(\lambda ^{ij} \tau^a_{\alpha \beta} N^a_i L_j^\alpha H^{\dagger \beta}  +
\frac{M_{N_i}}{2}   N^a_i N^a_i  +\hbox{h.c.})\,, 
\end{equation}
with $a=1,2,3$; $\alpha, \beta=1,2$.
As a result it leads to the same seesaw formula than with 
singlets: $M_\nu=-\lambda^T M_N^{-1} \lambda v^2$.
For leptogenesis, the one-loop diagrams are exactly the same as 
with right-handed neutrinos, Fig.~1, which lead to the same asymmetries up 
to $SU(2)_L$ factors of order unity,\cite{HLNPS}
\begin{equation}\label{epsT}
\varepsilon_{N_1}=\sum_{j=2,3}\frac{3}{2}
  \frac{M_{N_1} }{M_{N_j} }\frac{\Gamma_{N_j} }{M_{N_j} }
  I_j\frac{V_j-2 S_j }{3} \,,
\end{equation}
and is therefore 3 times smaller in the 
hierarchical limit where $V_j=S_j=1$, see Eq.~(\ref{eps}).
The final amount of baryon asymmetry is given by
the CP-asymmetry times the efficiency factor $\eta$ times a
numerical coefficient which is 3 times bigger than in the singlet case  
because now $N_1$ has three 
components: $\frac{n_{B}}{s} =-4.1 \cdot 
10^{-3} \varepsilon_{N_1} \eta$.
The decay width of each of the 3 components of $N_1$ 
is given by the same expression as in the singlet  
case, Eq.~(\ref{IGamma}).
Scatterings are same as in the type-I model except for $SU(2)_L$
factors\cite{HLNPS}.
The only important difference is that the triplets have 
$SU(2)_L \times U(1)$ gauge scatterings the singlets do not have.
This reduces the efficiency factor\cite{HLNPS}.
As a result, 
in the hierarchical limit $M_{N_1} << M_{N_{2,3}}$, 
the bounds for successful leptogenesis 
are slightly more stringent than for singlets:
\begin{equation}
M_{N_1} >1.5 \cdot 10^{10}~\hbox{GeV}\,, \,\,\,\,\,\,\,m_{\nu_3} < 0.12 
\hbox{eV} \,.
\label{boundfermions}
\end{equation}
In the quasi-degenerate case the discussion is similar to the one of 
the type-I model, see section 3.2.3 above. A value of $M_{N_1}$ as low as 
$\sim 1$~TeV is possible, very close to the resonance. 

\subsection{The Type-I plus Type-II Model}

The case where we add to the standard model three right-handed neutrinos 
and one scalar triplet is quite interesting because it is 
the situation of the ordinary left-right models and of the 
renormalizable $SO(10)$ models such as defined in Ref.\cite{Senj}.
In this case the relevant Lagrangian is just the sum of the Lagrangians 
of Eqs.~(\ref{Lseesaw}) 
and (\ref{Lseesaw2}).
To discuss this possibility it is necessary to consider two cases, 
depending on which particle is the lightest one, the scalar triplet 
$\Delta_L$ 
or the lightest right-handed neutrino $N_1$.

\subsubsection{The $M_\Delta < M_{N_1}$ case}

If the triplet is lighter than $N_1$, the production of the 
asymmetry will be naturally dominated 
by the decay of the triplet to two leptons.
The CP-asymmetry comes in this case 
from the difference between the decay width
of $\Delta_L^*$ to two leptons and of $\Delta_L$ to two anti-leptons.
The leptogenesis one loop diagram
is a vertex diagram involving both the decaying triplet 
and a virtual
right-handed neutrino,
first graph of Fig.~3. This diagram was first displayed 
in Ref.\cite{OS}.
Calculating explicitly its contribution we get\cite{HS}:
\begin{eqnarray}
\varepsilon_\Delta &=& 2\cdot
\frac{\Gamma (\Delta_L^* \rightarrow l + l) - \Gamma (\Delta_L \rightarrow 
\bar{l} + \bar{l })}{\Gamma_{\Delta_L^*}  
    +\Gamma_{\Delta_L}}\\
&=&  -\frac{1}{8 \pi}
\frac{M^2_\Delta}{ (\sum_{ij} |(Y_\Delta)_{ij}|^2 M^2_\Delta +|\mu|^2)} 
\frac{1}{v^2}
Im[(M^{I\ast}_\nu)_{il} (Y_\Delta)_{il} \mu^*]
\,,
\label{epsD}
\end{eqnarray}
where $M_\nu^{I}$ is the type-I contribution to the neutrino mass 
matrix (given in section 2). For each of 
the 3 triplet components the total decay width is:
\begin{equation}
\Gamma_\Delta=\frac{1}{8 \pi} M_\Delta \Big( \sum_{ij}|(Y_\Delta)_{ij}|^2 +
\frac{|\mu|^2}{M^2_\Delta} \Big) \,.
\label{gammaD}
\end{equation}
%
\begin{figure}[t]
\begin{center}
\begin{picture}(310,60)(0,0)
\DashArrowLine(0,30)(30,30){5}
\ArrowLine(60,60)(90,60)
\ArrowLine(60,0)(90,0)
\Line(60,60)(60,0)
\DashArrowLine(30,30)(60,60){5}
\DashArrowLine(30,30)(60,0){5}
\Text(3,22)[]{$\Delta_L^*$}
\Text(35,12)[]{$H$}
\Text(37,51)[]{$H$}
\Text(69,30)[]{$N_k$}
\Text(85,52)[]{$l_i$}
\Text(85,8)[]{$l_{l}$}
\Line(110,30)(140,30)
\DashArrowLine(170,60)(200,60){5}
\ArrowLine(170,0)(200,0)
\DashArrowLine(170,0)(170,60){5}
\DashArrowLine(140,30)(170,60){5}
\ArrowLine(170,0)(140,30)
\Text(115,22)[]{$N_k$}
\Text(145,12)[]{$l_l$}
\Text(147,51)[]{$H$}
\Text(182,30)[]{$\Delta_L$}
\Text(195,52)[]{$H^\ast$}
\Text(195,8)[]{$l_{i}$}
\DashArrowLine(210,30)(230,30){5}
\DashArrowArc(245,30)(15,180,360){5}
\DashArrowArcn(245,30)(15,180,0){5}
\DashArrowLine(260,30)(280,30){5}
\ArrowLine(280,30)(310,60)
\ArrowLine(280,30)(310,0)
\Text(215,22)[]{$\Delta^*_{L_i}$}
\Text(245,6)[]{$H$}
\Text(246,55)[]{$H$}
\Text(272,22)[]{$\Delta^*_{L_j}$}
\Text(297,55)[]{$l_{k}$}
\Text(297,4)[]{$ l_{l}$}
\end{picture}
\end{center}
\caption{One-loop diagrams contributing to the asymmetry
from the $\Delta_L$ and $N$ decays.}
\label{fig3}
\end{figure}
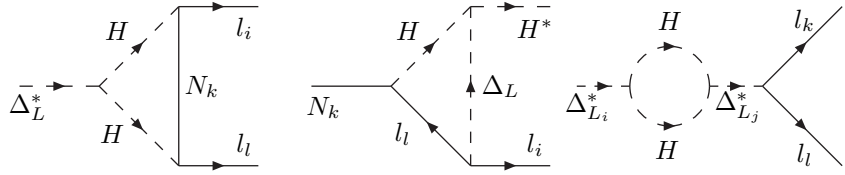
In this case leptogenesis works in a way similar to the type-I model 
apart from 3 important differences. The first is that, 
like for triplet of fermions above, there are 
gauge scatterings which tend to put the $\Delta_L$ closer to 
thermal equilibrium. The effect is similar to the one encountered for 
triplet of fermions above. The second difference is that here there is no 
one-loop self-energy diagram, so there is no possible 
resonance effect. The third one 
is that the interplay between the neutrino 
masses and the size of the washout and the size of the asymmetry 
is completely different from the type-I case.
The decay width is 
proportional to the triplet couplings but the asymmetry is essentially 
proportional to the right-handed neutrino couplings (taking into 
account the fact that there are two triplet couplings in both 
numerator and denominator of the asymmetry). As a result one may 
consider the possibility that the type-II contribution to neutrino masses is 
small enough to avoid large washout and that the type-I 
contribution to neutrino masses is large. By 
increasing in this way the type-I neutrino mass contribution, 
the washout remains unchanged but the asymmetry increases
proportionally to the neutrino masses\cite{HS}. Therefore, there is
no upper bound on neutrino masses in this model to have successful 
leptogenesis\cite{HS}. Note that 
there is nevertheless a lower bound 
on the triplet mass because 
the asymmetry is proportional to this mass.
For a hierarchical spectrum of light 
neutrinos (i.e.~$m_{\nu_3} \sim 
\sqrt{\Delta m^2_{\hbox{\footnotesize atm}}}$) the
 bound is essentially the same$^d$ as for the triplet of 
fermions, Eq.~(\ref{boundfermions}).
As there is no possible resonance effect
this bound is an absolute bound. For larger values of $m_{\nu_3}$ this 
bound decreases because in this case the asymmetry increases. 
However, assuming $m_{\nu_3}$ below 1 eV, this bound cannot 
decrease by more than one order of magnitude.  

\subsubsection{The $M_\Delta > M_{N_1}$ case}

If the triplet is sizeably 
heavier than at least one right-handed neutrino, then it is the decay 
of the right-handed neutrino to a lepton 
and a Higgs boson which dominates the production of the L asymmetry.
In this case leptogenesis can be produced from the pure type-I model 
just as in sections 2 and 3.
However here, in addition to this pure type-I contribution, there is 
a new contribution to leptogenesis coming from a 
diagram\cite{OS,laz1,HS,AK} with a real 
right-handed neutrino and a virtual triplet, second diagram of Fig.~3.
Disregarding the pure type-I contribution (assuming that it has a small 
contribution because the type-I Yukawa couplings and/or 
their phases are small)
this new diagram can perfectly lead to successful leptogenesis alone.
For the asymmetry defined in Eq.~(\ref{eps}) one obtains from this 
diagram\cite{HS,AK}
\begin{equation}
\varepsilon_{N_1}^\Delta=\frac{3}{16 \pi} \frac{M_{N_1}}{v^2} 
\frac{\sum_{il} Im[\lambda_{1i} \lambda_{1l}(M^{II*}_\nu)_{il}]}
{\sum_i |\lambda_{1i}|^2} \,,
\label{epsNDbis}
\end{equation}
where $M_\nu^{II}$ is the neutrino mass matrix induced by the type-II
contribution (given in section 4.1 above).
The discussion is similar to the one of 
the  $M_\Delta < M_{N_1}$ case above inverting
the role of type-I and type-II two contributions.
It is now the decay of the right-handed neutrino to lepton and Higgs,
induced by the type-I couplings, Eq.~(\ref{IGamma}), which essentially 
determines 
the washout. To increase the asymmetry without increasing 
the washout, one can then consider the possibility of keeping 
the type-I contribution small, increasing
the type-II contribution which increases the asymmetry but not the 
washout.
As a result, here too, there is no more upper bound on neutrino 
masses for leptogenesis\cite{HS}.
For hierarchical light neutrinos the lower 
bound on the lightest right-handed neutrino mass is to a good approximation
the same as 
in the pure type-I model\cite{HS}, Eq.~(\ref{MN1bound}). 
As the asymmetry is linear in both $M_{N_1}$ and the neutrino masses,
for larger values
of $m_{\nu_3}$ the asymmetry linearly increases,
so the bound on $M_{N_1}$ linearly decreases. But here too 
assuming $m_{\nu_3}$ below 1 eV,  
this bound cannot decrease by more than one order of magnitude.
The precise bound is given in Ref.\cite{AK} as a function of the 
efficiency factor $\eta$.

Note that, if instead of considering a dominant type-II 
contribution, we consider the case where both type-I and type-II contributions
are important this lower bound on $M_{N_1}$ doesn't get relaxed.
In this case both contributions 
appear to be just proportional to their respective 
contributions to the neutrino masses so that, baring
a possible cancellation of CP-violating phases, leptogenesis 
is expected to be dominated by the contribution which dominates the 
neutrino masses\cite{HS}.
From Eq.~(\ref{eps}) the pure type-I contribution to leptogenesis 
for hierarchical right-handed neutrinos turns out to be 
given by Eq.~(\ref{epsNDbis}) replacing $M_{\nu}^{II}$ by 
$M_{\nu}^I$, a fact which can be nicely understood by 
using effective dimension five neutrino mass operators.\cite{AK}

For a quasi-degenerate spectrum of right-handed neutrino the pure type-I
contribution to leptogenesis is expected dominant and the triplet 
contribution which doesn't display any resonant behaviour is in this case
negligible. 

\subsection{The Multiple Type-II Models}

If there is more than one heavy scalar triplet, leptogenesis can be 
easily induced by the decay of the triplets to two leptons with
a one-loop self-energy diagram involving two different 
triplets\cite{MS1,HMS1},
third diagram of Fig.3.
The asymmetry in this case is given by\cite{MS1}
\begin{equation}
\varepsilon_{\Delta_i} =  -\frac{1}{ \pi} M_{\Delta_i}
\frac{ Im[(\mu^*_i \mu_j (Y_{\Delta_i})_{kl}
 (Y^*_{\Delta_j})_{kl} ] }{  |(Y_{\Delta_i})_{kl}|^2 M^2_{\Delta_i} 
+|\mu_i|^2}  \frac{M^2_{\Delta_j}  
\Delta M^2_{1j}}{(\Delta M^2_{1j})^2+M_{\Delta_1} ^2
   \Gamma_{\Delta_j} ^2}
\,,
\label{epsDi}
\end{equation}
where there is now a triplet scalar index on the couplings 
of Eqs.~(\ref{Lseesaw2}) and (\ref{gammaD}). 
For hierarchical triplets, under the assumption of footnote d above,  
successful leptogenesis leads to a triplet mass bound similar to
the one of Eq.~(\ref{boundfermions}) for hierarchical fermion triplets,
in accordance with the estimate of Ref.\cite{HMS1}.
This bound is higher than 
for right-handed neutrinos due to gauge scatterings. Similarly for 
quasi-degenerate triplets, one can go down to 1-10 TeV (very close to 
the resonance). And,
here too, even with hierarchical triplets, 
there is no more upper bound on neutrino masses for successful 
leptogenesis because one can always increase the neutrino mass contribution
of the virtual triplet in the self-energy diagram leading to a 
larger asymmetry without increasing the washout.

\subsection{The Multiple Type-I Seesaw Case}

To add extra fermion singlets $S_i$ on top of the 3 right-handed 
neutrinos $N_i$ is 
also a possibility one might consider,\cite{AB,Ut,AAL} especially 
if these extra 
$SU(2)_L \times U(1)$ singlets are 
also singlets of SO(10) in case one can build a 
renormalizable SO(10) model\cite{AB,Ut} without a 126 
scalar multiplet to give mass to the $N_i$'s.
In this case, as there is no 126, there is no triplet.
This may lead to successful leptogenesis just from the same diagrams as 
in the type-I model, Fig.~1, 
except that there are now more than 3 singlets to be 
put in these diagrams. There are diagrams where both 
real and virtual singlets are $N_i$'s or are $S_i$'s.
Clearly, for example if the lightest heavy particle is a $S$, one can 
increase the Yukawa couplings of the $N$ leaving the $S$ Yukawa 
couplings unchanged. In this way the neutrino masses increase but not 
the washout which is due the $S$ couplings, so that, 
here too, there is no relevant upper bound on neutrino masses for successful 
leptogenesis.

Note that extra diagrams with a real $N$ and a 
virtual $S$ in Fig.1 (or viceversa)
are also possible if they couple to the same scalar multiplets.
In the SO(10) models of  Ref.\cite{AB,Ut} the $S$ and $N$ do not 
couple to the same multiplets because the $S$ which is a singlet of SO(10)
couples to the 16 of matter $\psi_{16}$ and a scalar 
$H_{\overline{16}}$ 
but the $N$ which is in $\psi_{16}$ couples to an other $\psi_{16}$ 
and a scalar $H_{10}$.
However through mixing of the $H_{\overline{16}}$ 
and $H_{10}$ from a coupling 
involving the vev of another $H_{\overline{16}}$ such a diagram may exist, a 
possibility still to consider.
It is straightforward to check that such a diagram can lead to 
successful leptogenesis and here 
too without leading to a relevant upper bound on the neutrino masses.

Note also that there is an other possibility with extra singlets\cite{AAL}, 
simply by assuming
more than three generations of fermions and by assuming a huge hierarchy of 
Yukawa couplings (to have $m_{\nu_4}> 45$~GeV). This may also lead to 
successful leptogenesis (e.g.~at a low scale).

\section{The Radiative Models}

An other class of models one might consider to explain the neutrino 
masses and mixings is the one where neutrino masses are induced by 
loop diagrams. This is quite interesting as in most of these models 
the scale where they are generated is not far beyond the reach of
present particle accelerators.
However to generate leptogenesis at the $\sim$~TeV scale in this framework 
turns out to be a quite hard task for various reasons.
A first reason is that in these models, such as the ones based on 
violation of R-parity\cite{Rparity} or 
based on the presence of a charged scalar 
$SU(2)_L$ singlet (Zee model\cite{Zee}), the 
heavy states whose decays could generate the 
lepton asymmetry are not gauge singlets. As a result there is a large 
washout suppression coming from gauge scatterings involving these 
heavy states which are very fast at low scale. A second reason is that 
at the TeV scale the Hubble constant is much smaller than for example 
at $10^{10}$~GeV
since it is proportional to the square of the temperature.
Since the decay width is only linear in the mass, the condition $\Gamma < H$
requires therefore couplings much smaller at the TeV scale than 
at $10^{10}$~GeV. 
Since the asymmetry is proportional to these tiny 
couplings, this leads in general to too small 
lepton asymmetry, see Ref.\cite{TH}. 

One solution to these problems is to consider 
three body decays instead of two-body decays.\cite{TH}
An other solution is to consider
the seesaw extended MSSM, that is the say the MSSM extended by 
three right-handed (s)neutrinos. In this framework 
it has been shown\cite{BHS} that a large enough asymmetry can be obtained 
from the L-violating soft 
supersymmetry breaking terms involving the right-handed sneutrinos.
With sneutrino masses of order a few TeV, successful
leptogenesis together with light neutrino masses (induced 
radiatively from the same soft terms) can be induced.

\section{Summary}

In summary there are quite a few models which from the same interactions 
can lead to successful generation of neutrino masses and leptogenesis:
type-I model with right-handed neutrinos, type-II model (although only if 
involving the soft terms), type-III model with triplet of 
fermions, type-I plus type-II model, multiple type-I or type-II models, 
or even a radiative model (in the seesaw extended MSSM from soft terms).

The upper bound on neutrino masses is quite sensitive to the model 
considered as well as on the assumptions made on the heavy mass spectrum 
in each model.
In the type-I model (and similarly in the type-III model) 
a stringent bound, Eq.~(\ref{mnu3bound}), can 
be found only 
assuming a hierarchical spectrum of right-handed neutrinos. However
this assumption
doesn't appear 
to be the most natural assumption one can make 
when considering this bound, for which
the light neutrinos have a quasi-degenerate 
spectrum. A probably more natural 
quasi-degenerate spectrum of right-handed neutrinos
leads instead in full generality to an upper bound on neutrino masses 
far beyond the eV scale. Even in a highly constrained model, such as the 
one based on 
a SO(3) 
symmetry in section 3.2.3, 
a value around 1 eV appears to be possible for successful 
leptogenesis.
Only with extra assumptions (such as low reheating temperature) one might 
get a more stringent bound.
In the other models (multiple type-I and/or type-II) the upper bound 
is far above the eV scale even for a hierarchical spectrum of heavy states.

Similarly the lower bound on the scale of leptogenesis is sensitive to 
the model and/or the heavy mass spectrum considered. However for this 
scale there is at least one firm conclusion one can draw:
in all the seesaw models we considered here, if 
the masses of the heavy states differ by orders of magnitude,  
this scale has to be orders of magnitude above the 
TeV scale, that is to say 
above $10^7-10^{10}$ GeV depending on the model.

\section*{Acknowledgments}
It is a pleasure to thank L. Boubekeur, G. D'Ambrosio, A. Hektor,
Y. Lin, E. Ma, A. Notari, 
M. Papucci, M. Raidal, A. Rossi, 
U. Sarkar, G. Senjanovic and A. Strumia with whom part of the work presented 
here was done. We also thank A. Faraggi and U. Sarkar for discussions on 
multiple type-I models. This work was supported by the Marie Curie 
HPMF-CT-01765 and EU MRTN-CT-2004-503369 (The quest for unification) 
contracts.

\end{document}